\documentclass[aps,prb,twocolumn,showpacs,amsmath,amssymb,superscriptaddress, citeautoscript]{revtex4-2}
\setcitestyle{super}
\usepackage{graphicx}
\usepackage{hyperref}
\usepackage{xcolor}
\usepackage{color}
\usepackage[all,cmtip]{xy}
\usepackage{multirow}
\usepackage{natbib}
\usepackage{amsmath,amsthm} 
\usepackage{amsfonts}

\makeatother

\begin{document}

\title{On the average spin Chern number}

\author{Rafael Gonz\'{a}lez-Hern\'{a}ndez}
\email{rhernandezj@uninorte.edu.co}
\affiliation{Departamento de F\'{i}sica y Geociencias, Universidad del Norte, Km. 5 V\'{i}a Antigua Puerto Colombia, Barranquilla 081007, Colombia}
\affiliation{Institut f\"ur Physik, Johannes Gutenberg Universit\"at Mainz, D-55099 Mainz, Germany}
\author{Bernardo Uribe}
\email{bjongbloed@uninorte.edu.co}
\affiliation{Departamento de Matem\'{a}ticas y Estad\'{i}stica, Universidad del Norte, Km. 5 V\'{i}a Antigua Puerto Colombia, Barranquilla 081007, Colombia}

\date{\today}

\begin{abstract}
	 In this work we propose the average spin Chern number (ASCN) as an indicator of the topological
significance of the spin degree of freedom within insulating materials.
Whenever this number is a non-zero even integer, it distinguishes the  material as a spin Chern insulator, and the number is a topological invariant whenever there is a symmetry that commutes with the spin and protects Chern numbers. 
If this number is not zero, it indicates that the material has non-trivial spin transport properties, and it lies close to the value of the 
spin Hall conductivity (SHC) within the bandgap. 
For systems where the spin commutes with the Hamiltonian, the ASCN matches the SHC. 
When the non-commutativity of the spin with the Hamiltonian cannot be neglected, both values are non-zero simultaneously.
The ASCN is therefore a good complement for the intrinsic contribution of the SHC, and permits to detect
topological information of the material which is not possible alone from the value of the SHC.
\end{abstract}
\maketitle

\section*{Introduction}

The investigation of topological phases of matter holds a central position in the landscape of condensed matter physics, tracing its origins to breakthroughs such as the quantum Hall effect and quantum spin Hall effect \cite{TKNN-invariant,Haldane-model,sinova2015spin,kane2005quantum}. 
The discovery of a constant value within the band gap of the spin Hall conductivity (SHC) in topological insulators presents a compelling challenge, stimulating significant interest and inquiry \cite{Bernevig2006,Fukui07,qi2008topological,hasan2010colloquium,qi2010topological}.
While significant progress has been made in elucidating this issue for 2D topological insulators, where the value becomes quantized in accordance with the definition of the spin Chern number (SCN), this quantization is applicable only in the non-relativistic limit or weak spin-orbit coupling (SOC) interactions \cite{Quantized-QSHE,SCN-liev,SHE-in-2d-TI,Ezawa14,Measurement-SCN,single-point-SCN}. 
Nevertheless, a critical question remains unresolved, particularly concerning 3D topological insulators \cite{SHE-in-3d-TI}.
The constant value of the SHC within the band gap eludes a comprehensive explanation, even in the non-relativistic limit.

The conventional approach to topological features of matter typically focuses on the exploration of electronic band structures \cite{band-topology,Topological-quantum-chemistry}.
However, when considering the incorporation of spin quantum degree of freedom, an additional step is required: the projection of spin operator onto the valence electronic states \cite{Prodan-SCN,spin-resolved-3d,Spin-texture-fragile}. 
This process produces the spin spectrum, with the intrinsic topology originating from the nontrivial SCN. 
Using this approach, Prodan \cite{Prodan-SCN} defined the SCN in the thermodynamic limit, demonstrating its robustness against disorder and smooth deformations of the Hamiltonian.
 Fu  \textit{et al}\cite{SCN-optical-lattices} extended the concept to the real-space SCN for optical near fields of finite-sized structures.  
 Recently, Lin \textit{et al}\cite{spin-resolved-3d} introduced the nested spin-resolved Wilson loops and layer constructions as tools to characterize the intrinsic bulk topological properties of 3D spinful insulators. Particularly focusing on identifying three distinct spin-resolved phases in helical higher-order topological insulators: 3D quantum ulators, spin-Weyl semimetals, and T-doubled axion insulator states. 
This spin topology classification approach provides robust topological invariants for studying the spin transport response of quantum systems \cite{spin-resolved-3d,Gonzalez-Uribe}.

In this study, following the definition of Prodan \cite{Prodan-SCN} of the spin Chern number,
we introduce the concept of average spin Chern number (ASCN) as a strategic tool to unravel the nature of SHC within the bandgap in 3D topological insulators.
We show that the ASCN may offer valuable insights for the information underlying the SHC in insulator materials,
thus indicating the interplay between spin properties and topological features in such systems. 
By exploring the connection between SHC and ASCN, our work aims to contribute to a deeper understanding of spin transport phenomena in 3D topological insulators. 
This work can be seen as a continuation of our previous efforts to understand
spin properties in materials\cite{Gonzalez-Uribe}. It is based on many sources, but in particular it
is framed in the ideas and constructions of the colleagues working on spin-resolved topology \cite{spin-resolved-3d}.

\section{Projected spin topology}

In an insulator, the valence states form a complex vector bundle that is known by the name of {\it Bloch bundle}. The Bloch bundle is endowed with the action  of the group of crystal symmetries and moreover,
it may posses time reversal symmetry,  or a composition of time reversal symmetry with a rotation, inversion or a translation in the magnetic case \cite{space-matter-topology}. Incorporating spin-orbit interactions implies that the Bloch bundle
is endowed with the action of the lift of the crystal symmetries to the spin group, and that time reversal operator squares to $-1$.

All these symmetries endow the Bloch bundle with a rich variety of topological invariants, some of them related to magneto-electric properties of the crystal \cite{qi2008topological,Magnetoelectric-Vanderbilt}. Among those topological invariants we may mention
the Chern-Simons axion coupling term and the Chern classes  \cite{Vanderbilt-axion,moore_book}. The first one is equivalent to the Kane-Mele
invariant in the presence of time reversal symmetry \cite{Fu-Kane,Fu-Kane-Mele}, and this invariant gives the indication of a strong topological insulating phase.

When taking into consideration the spin operator $S_n$ in any direction $n$ (here $n$ is a unit vector in $\mathbb{R}^3$), we see that the Bloch bundle is moreover endowed with the action induced by this operator once it has been projected to the valence states. In other words, if $P$  is the projection operator into the valence states
\begin{align}
P := \sum_{n \in \mathrm{occ}} |\psi_n \rangle \langle \psi_n |
\end{align}
where $\psi_n$ are the eigenstates of the Hamiltonian and $occ$ denotes the occupied states of an insulator, the operator $P S_ n P$ defines a Hermitian operator on the Bloch bundle
and its eigenvalues are between $-1$ and $1$ (in $\hbar/2$ units) \cite{Prodan-SCN}.
The spectrum of this projected operator will define bands whose eigenvalues are
between $-1$ and $1$, and the separation into positive and negative bands permits
one to obtain topological information on the bundle.

  On Hamiltonians with time reversal symmetry the positive eigenvectors are mapped to the negative ones by the time reversal operator, thus implying that
 the projected spin spectrum is symmetric with respect to zero. The projected spin operator separates the valence states into
 positive and the negative eigenvectors except in those points where the projected spin spectrum contains the zero. These points are localized where the positive
 bands of the projected spin operator cross the negative ones, and are called spin Weyl points (SWP). Their presence or absence have direct topological consequences within the material; particularly evident is the spin transport response which we have explored in our previous work \cite{Gonzalez-Uribe}.

The projected spin operator may be gapped, thus separating positive from negative bands
in the Bloch bundle, or it has only a finite number of SWPs in the reciprocal space.
In both cases there are no SWPs in all but a finite number of $k$-planes. Hence the Chern numbers of the positive and negative eigenstates of the projected spin operator can be determined in all but finitely many $k$-planes.

 Denoting by $\{\phi_m^{+}| m \in \mathfrak{S}^+\}$ and  $\{\phi_m^{-}| m \in \mathfrak{S}^-\}$ the positive and negative  valence eigenvectors of the projected spin operator, and making a choice 
of oriented coordinate axes $abc$,
 the Chern numbers of the positive and negative eigenvectors on the plane $k_c=l$ with $l \in [0,2\pi]$, can be calculated as follows:
 \begin{align}
 c^{S_n,\pm}_{1,ab}(l) : = \tfrac{1}{2\pi} \int \! \! \! \int  \tfrac{dk_a dk_b}{(2\pi)^2}\sum_{m \in \mathfrak{S}^\pm}  2\  \mathrm{Im} \langle \tfrac{\partial \phi_m^{\pm}}{\partial k_a} | \tfrac{\partial \phi_m^{\pm}}{\partial k_b} \rangle .
 \end{align}
 
 Since there are only a finite number of SWPs, then these Chern numbers
are defined for all the planes $k_c=l$ except for a finite number of them where the SWPs lie. 
 
 The spin Chern number is thus:
\begin{equation}
c_{1,ab}^{S_n}(l) := c_{1,ab}^{S_n,+}(l) - c_{1,ab}^{S_n,-}(l),
\end{equation}
and it is a well defined integer for all but finite number of $k_c$-planes. For systems on which
the first Chern number is zero on all $k$-planes, the spin Chern number is always twice the value of the Chern number of the positive eigenvectors. This is the case on systems with time reversal symmetry.

Here we would like to emphasize that we are defining the spin Chern number as the difference between the Chern numbers of the positive and negative valence bands with respect to the projected spin operator. The original definition used by Kane and Mele divides this quantity by two\cite{kane2005quantum}.
We follow the definition of Sheng \textit{et al} \cite{Quantized-QSHE} and not divide by two since we would like the spin Chern number to also take integer values on magnetic materials. 

\subsection{Average spin Chern number}

We want to extract relevant topological information from the Bloch bundle using the 
whole range of spin $S_n$ operators and all possible planes $\{xy,yz,zx\}$. So, fixing a spin direction  $S_n$ and a plane direction $ab \in
\{xy,yz,zx\}$, we propose to consider
the {\it average spin Chern number} (ASCN) of the projected spin operator $PS_nP$ across all planes in reciprocal space $k_c=l, l\in [0,2\pi]$:
\begin{equation}  \label{definition ASCN}
\overline{c}_{1,ab}^{S_n} = \tfrac{1}{2\pi} \int_{0}^{2\pi} c_{1,ab}^{S_n}(l)  dl.
\end{equation}
For a set of axes with the opposite orientation $bac$ we define the ASCN as:
\begin{equation}
\overline{c}_{1,ba}^{S_n}  := -\overline{c}_{1,ab}^{S_n}.
\end{equation}

We emphasize that the ASCN depends on a choice of spin direction (or spin quantization axis) and a choice of $k$-planes on which the spin Chern numbers are calculated. Thus the ASCN should be understood as a family of quantities associated to the material. 
 Appropriate choices of the spin direction and of the plane of integration are important in order to gain some hindsight on the spin properties of a material.

The relevance of the ASCN is based on the following properties:

\begin{itemize}
\item[1.] {\it Whenever the spin operator commutes with the Hamiltonian, the intrinsic contribution of the SHC
is a multiple of the ASCN:
\begin{equation}
{\sigma}_{ab}^{c}  = - e \pi \cdot \overline{c}_{1,ab}^{S_c}.
\end{equation}
Here $abc$ is a base in a Cartesian coordinate system.}
\end{itemize}
It is known that whenever the spin is a well-defined quantity, the 2D
SHC is a multiple of the SCN\cite{Quantized-QSHE, QSHE-no-TRS, Prodan-SCN, SHE-in-2d-TI}.   
In 3D materials where the Hamiltonian commutes with the spin, the SCN is a multiple of the SHC in all but a finite number of $k$-planes. The SHC can be thus seen as the integral of the spin Chern numbers (SCNs) along parallel planes and therefore it is a multiple of the ASCN. 

To make this statement precise, choose an orthogonal basis $abc$ in the cartesian coordinate system. 
Since the spin $S_c$ and the Hamiltonian commute, one can choose a basis of states  $\psi_m^{\pm}$ which are eigenvectors of both operators:
  \begin{align}
  H \psi_m^{\pm} = E^{\pm}_m \psi_m^{\pm} \ \ \mathrm{and} \ \ \  S_c \psi_m^{\pm} = \pm \psi_m^{\pm}.
  \end{align}
 The SCN in the plane $k_c=l$ can be written as
  \begin{align}
 c^{S_c}_{1,ab}(l) = \tfrac{1}{2\pi} \int \! \! \! \int  \tfrac{dk_a dk_b}{(2\pi)^2}\sum_{m \in \mathrm{occ}}  2\  \mathrm{Im} \langle \tfrac{\partial (S_c \psi_m)}{\partial k_a} | \tfrac{\partial \psi_m}{\partial k_b} \rangle 
 \end{align}
 where $m$ runs over the valence states (occupied) and $S_c \psi_m = \pm \psi_m$ picks up the negative sign for 
 the Chern number of the negative projected spin eigenvectors.
 
 Adding the term $|\psi_r\rangle \langle \psi_r|$ running over all states (identity operator), we get the term
 \begin{align} \label{term with identity}
 \langle \tfrac{\partial (S_c \psi_m)}{\partial k_a} |\psi_r\rangle \langle \psi_r| \tfrac{\partial \psi_m}{\partial k_b} \rangle.
 \end{align}
Using the identity 
\begin{align}
(E_m-E_r) \langle \psi_r | \nabla_{\mathbf{k}} \psi_m \rangle =  \langle \psi_r | \nabla_{\mathbf{k}}H | \psi_m \rangle
\end{align}
we see that the term of Eqn. \eqref{term with identity} becomes 
\begin{align}
\frac{\langle \psi_m | H_a S_c | \psi_r \rangle \langle  \psi_r |H_b | \psi_m \rangle }{(E_m-E_r)^2}
\end{align}
 where $H_a$ denotes $\tfrac{\partial H}{\partial k_a}$ and $H_b$ denotes $\tfrac{\partial H}{\partial k_b}$.
 Replacing $H_a S_c$ by $\tfrac{1}{2} \{H_a,S_c\}$ and noting that
  $\langle \tfrac{\partial \psi_m}{\partial k_a} | \psi_m \rangle$ is real, we get the following formula for the SCN
  on the plane $k_c=l$:
  \begin{align}
 c^{S_c}_{1,ab}(l) = &  \int \! \! \! \int  \tfrac{dk_a dk_b}{(2\pi)^3}  \\
&  \sum_{m \in \mathrm{occ}}
 \sum_{r \neq m}  2\  \mathrm{Im} \tfrac{\langle \psi_m | \tfrac{1}{2} \{H_a,S_c\} | \psi_r \rangle \langle  \psi_r |H_b | \psi_m \rangle }{(E_m-E_r)^2}.  \nonumber
 \end{align}
 Further replacing the spin current operator $\hat{j}^c_a = \tfrac{1}{4}\{S_c, H_a\}$ and the velocity operator
 $\hat{v}_b= \tfrac{1}{\hbar} H_b$ we get:
  \begin{align}
 -\tfrac{e}{2} c^{S_c}_{1,ab}(l) = &  \hbar \int \! \! \! \int  \tfrac{dk_a dk_b}{(2\pi)^3}  \\
&  \sum_{m \in \mathrm{occ}}
 \sum_{r \neq m}  2\  \mathrm{Im} \tfrac{\langle \psi_m |\hat{j}^c_a  | \psi_r \rangle \langle  \psi_r |-e \hat{v}_b | \psi_m \rangle }{(E_m-E_r)^2}.  \nonumber
 \end{align}
 The Kubo formula for the intrinsic contribution of the SHC is \cite{Calculation_intrinsic_SHC}:
  \begin{align}
\sigma_{ab}^c = &  \hbar\int dk_c \int \! \! \! \int  \tfrac{dk_a dk_b}{(2\pi)^3}  \label{Kubo SHC} \\
&  \sum_{m \in \mathrm{occ}}
 \sum_{r \neq m}  2\  \mathrm{Im} \tfrac{\langle \psi_m |\hat{j}^c_a  | \psi_r \rangle \langle  \psi_r |-e \hat{v}_b | \psi_m \rangle }{(E_m-E_r)^2},  \nonumber
 \end{align}
 and therefore we obtain the desired equality
 \begin{align}
 \sigma_{ab}^c = \int_0^{2 \pi} -\tfrac{e}{2} c^{S_c}_{1,ab}(l) dl = - e \pi \cdot \overline{c}^{S_c}_{1,ab}.
 \end{align}
 
 Here we point out that whenever the spin operator does not commute with the Hamiltonian,
 the 2D SHC on $k$-planes is not a quantized quantity, as demonstrated by various authors \cite{Quantized-QSHE, QSHE-no-TRS, SHE-in-2d-TI,single-point-SCN}
 On the other hand, the SCN is a clearly defined even integer number. This is the main reason we chose to
define its average on the 3D Brillouin zone.

 \begin{itemize}
\item[2.] {\it The ASCN is not-linear on the spin direction $n$. }
\end{itemize}
The ASCN is not linear on the vector $n$ defining the spin $S_n$, in contrast with the intrinsic contribution of the SHC which is linear by definition.
 Topological invariants such as the SCN cannot vary linearly on $n$ since its values are integers. Take for example the spin Chern number $c^{S_n}_{1,xy}(0)$ across  the plane $k_z=0$ and
 vary the spin $S_n$ with $n=(\cos(\theta),0,\sin(\theta))$, $0 \leq \theta \leq \pi$. 
 The spin Chern number $c^{S_n}_{1,xy}(0)$ as a function of the variable $\theta$
 is a step function with even integer values. Hence not-linear in the spin direction.
 
\begin{figure}
 	\includegraphics[width=8.9cm]{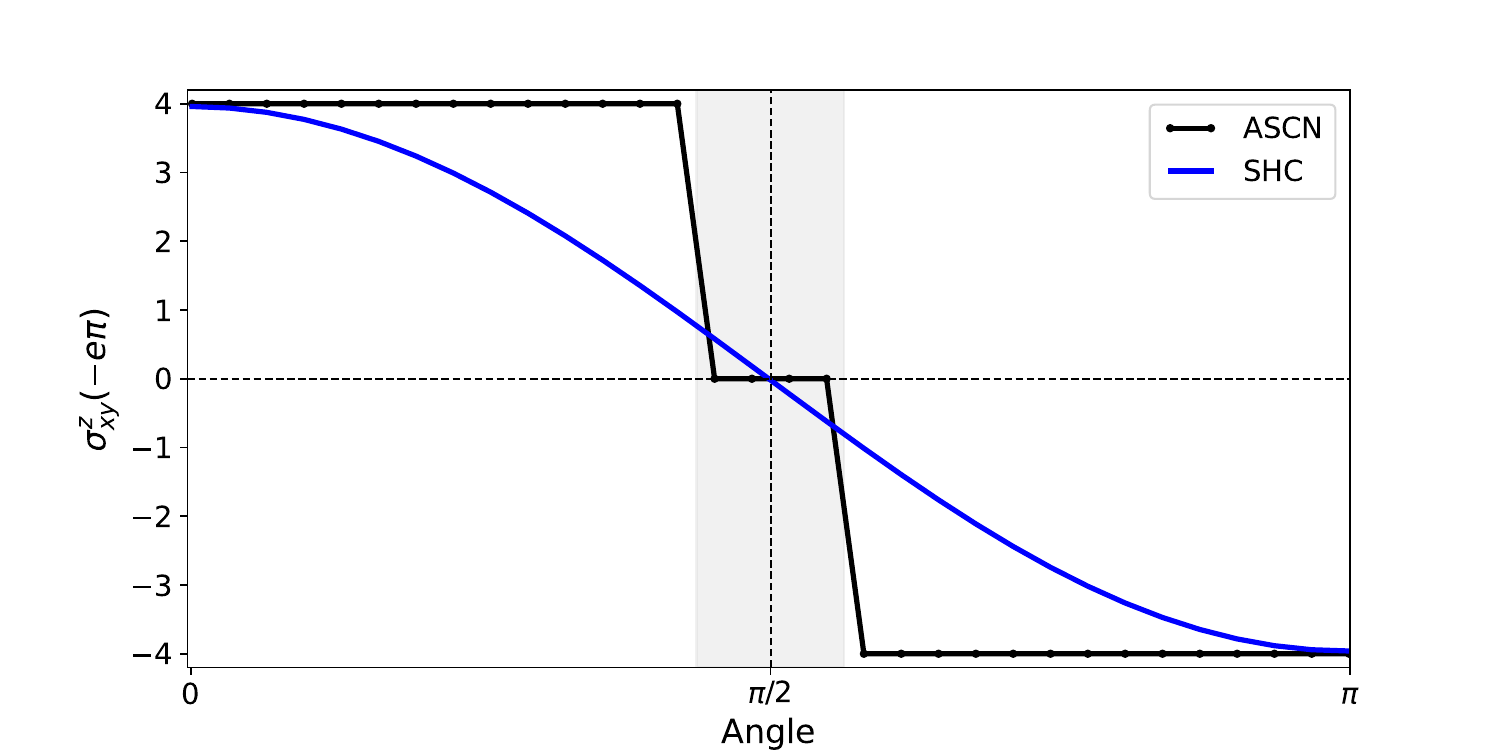}
 	\caption{SHC $n \cdot (\sigma_{xy}^x,\sigma_{xy}^y,\sigma_{xy}^z)$ (blue lines) and ASCN $\overline{c}_{1,xy}^{S_n}$ (black line with dots) for the $\alpha$-BiBr material. Spin $S_n$ is defined with $n=(\sin(\theta), 0,\cos(\theta))$ where $0 \leq \theta \leq \pi$.  The transition is presented from $S_z$ to $S_x$ and finish in $S_{-z}$. Quantum spin Hall insulator behavior is noted in proximity to the spin directions $S_z$ and $S_{-z}$.  Note that there are three different phases for the ASCN, and that on each change of phase, the projected spin gap closes. On the spin directions where the ASCN is zero there is not topological contribution of the spin properties on the planes parallel to $k_xk_y$.}
 	\label{bibr}
 \end{figure}

\begin{figure}
	\includegraphics[width=8.9cm]{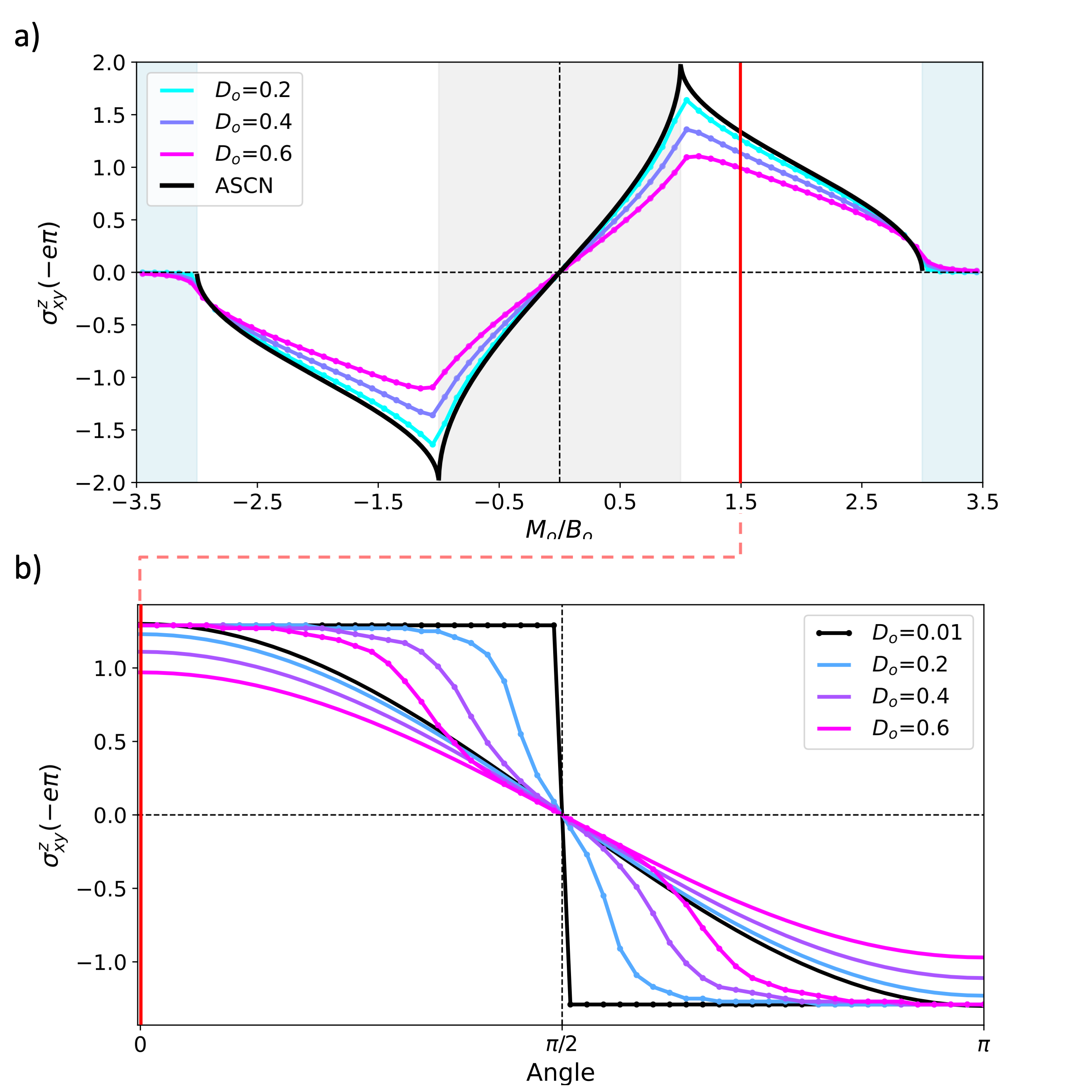}
\caption{Comparison of SHC with the ASCN in the 3D BHZ model using various computational and theoretical approaches. a) Results obtained from the Kubo formula \eqref{Kubo SHC} of the SHC $\sigma_{xy}^z$ for different
values of $D_0$ (in colors), and the ASCN  $\overline{c}_{1,xy}^{S_z}$ of \eqref{definition ASCN} in reciprocal space (black line),  are plotted against the $\tfrac{M_0}{B_0}$ parameters. 
The shaded areas represent different phases: the blue region denotes the trivial phase, while the white and gray areas indicate the strong and fragile phases, respectively. b) Comparison
of the SHC  $n \cdot (\sigma_{xy}^x,\sigma_{xy}^y,\sigma_{xy}^z)$ (lines without dots) with the ASCN
 $\overline{c}_{1,xy}^{S_n}$ (lines with dots) for the spin $S_n$ where
$n=(\sin(\theta), 0,\cos(\theta))$ and $0 \leq \theta \leq \pi$. Here we go from $S_z$ to $S_x$ and finish in $S_{-z}$. 
The graph a) is at angle $\theta =0$ and the graph b) is at $\tfrac{M_0}{B_0}=1.5$. The red vertical line
in the graphs represent the same information.
  The strength of spin-orbit coupling ($D_0$) is measured in electron volts (eV). }
	\label{shcvsmobo}
\end{figure}

  This non-linearity can be seen in Fig. \ref{bibr} where the ASCN and the SHC have been calculated for $S_n$ varying from $S_z$ to $S_{-z}$  across the planes parallel to $xy$ in the material $\alpha$-BiBr. In this material the projected spin spectrum is gapped, so the 
 ASCN is a step function with even integer values.
   Also in
  Fig. \ref{shcvsmobo} b) the ASCN and the SHC have been calculated for different values of $n$ and for different coefficients in the 3D BHZ model.
  On this model, whenever the coefficient $D_0$ goes to zero, the location of the SWPs become less  dependent of the variable $\theta$. This is why the ASCN looks like a 
  step function for the case $D_0=0.01$. The non-linearity of the ASCN with respect
  to the spin direction is evident.

\begin{itemize}
\item[3.] {\it If the ASCN is zero in all spin directions and all reciprocal planes then there are no SWPs
 and all spin Chern numbers vanish. }
\end{itemize}
The location of the SWPs is sensitive to the direction of the spin. This can be observed
in the calculation carried out in \S\ref{subsection SWP for generic spin} for the position of the SWPs for a generic spin in the 3D BHZ model presented in \S\ref{section 3D BHZ model}. Therefore, if the ASCN is uniformly zero in all directions, it indicates the absence of SWPs. Any presence of SWPs in any spin orientation would result in a non-zero ASCN value.

  Therefore, not only there are no SWPs, but also the SCNs remain constant and equal to zero.
Thus the positive eigenvectors of the projected spin operator can be separated from the negative ones. The Bloch bundle is hence a sum of positive and negative bundles, and since
 the Chern numbers of both bundles are zero across all planes, there is no contribution of the projected spin operator on the SHC. The system is trivial in terms of Chern numbers
 and SHC, but it may still have topological significance. The positive (and negative) bands might have non-trivial Chern-Simons $\theta$-term, potentially resulting in non-trivial surface SHC for each band group\cite{Vanderbilt-axion}.

\begin{itemize}
\item[4.] {\it If the ASCN is an even integer for generic directions of the spin, it means that the projected spin operator has no SWPs. Therefore the material
is a spin Chern insulator in the spin directions where the ASCN is non-zero. Moreover,
if there is a symmetry that commutes with the spin and protects the Chern numbers,
then the ASCN is a topological invariant.}
\end{itemize}

Whenever time reversal symmetry is present, the SCN is always an even integer. This
follows from the fact that in the planes $k_c=0$ the SCN is twice the value of the 
Chern number of the positive bands. The value of the SCN only changes by multiples of two whenever the plane is moved to $k_c=l$ and the value changes whenever SWPs are crossed.

If the ASCN is an even integer for generic directions of the spin, it means
that the ASCN is constant for directions of the spin that are close to one another, and therefore there cannot be SWPs. We show in  \S\ref{subsection SWP for generic spin}
that the location of the SWPs is sensitive to the direction of the spin, and since the ASCN
is constant, there cannot be SWPs. If there are no SWPs for a choice of spin direction, then the SCNs are equal on parallel planes. 

A {\it spin $S_n$ Chern insulator} (SCI) is an insulator with gapped projected spin
operator $PS_nP$ whose positive bands (also negative) carry a non-trivial first Chern number \cite{Gonzalez-Uribe}. The value of the ASCN parallel to $k_c=l$ planes is exactly twice the value of the Chern number of the positive bands. 

Changing the direction of the spin does not close the projected spin gaps unless there is a change of topological phase for the spin operator. 
The change of direction of the spin might change the Chern number for the positive and negative eigenvectors of the projected spin operator, and when this happens, there is a change in the topological phase. This change of phase happens in a set of dimension 1 (curve) inside the sphere of all possible directions for the spin. Whenever there is a change
of phase we claim that the ASCN changes by multiples of $4$. The reasoning is the following.

In the generically gapped projected spin operator, the ASCN is a constant even number.
Since ASCN is locally constant, then the SCNs are also locally constant on each plane $k_c=l$.
For the ASCN to change, all the SCN must change simultaneously. This means that there
must be zeros of the projected spin operator in all planes $k_c=l$ for a specific choice of 
spin direction $S_n$. 
These crossings must come in pairs for each plane $k_c=l$ since
they appear twice on the plane $k_c=0$ due to the time reversal operator, and the change
in SCN is the same across all parallel planes. Since on each plane, we have two crossings
of the projected spin operator, the change in SCN is by multiples of $4$. Hence the change
of phase for the ASCN is also by multiples of $4$

Whenever the ASCN is an  even integer, we also might expect this number to be a topological invariant of the system. This is indeed the case if there a symmetry on the material
commuting with the spin operator that moreover protects the Chern numbers on the positive and the negative bands.   This is for example the case whenever inversion is a symmetry of the material. Inversion commutes with the spin operators, and protects the Chern numbers
on the planes $k_c=0,\pi$. But whenever there is no such symmetry commuting with the spin that protects the Chern numbers, then there is no reason to expect that the local SCNs are protected. This is the case of the SCN in 2D materials in the presence only of time reversal symmetry. The topological invariant associated to the SCN in this case is only its value modulo 4\cite{Prodan-SCN}.

The even integer value of the ASCN
indicates that the material has a constant SCN across all parallel $k$-planes in a fixed spin direction. This number does not change
on adiabatic deformations of the Hamiltonian if there is a symmetry that commutes with the spin and
which moreover protects Chern numbers.
This particular phase is also recognized as the 3D quantum spin Hall insulator state, in analogy with the 2D\cite{Bay-antimonene,QSHI,QSHIAFM,BiBr_exp2022,BiBr_Yao}.
 
The previous phenomena can be seen in Fig. \ref{bibr} where the material $\alpha$-BiBr has been studied for different directions of the spin. Here the projected spin operators are gapped,
 the ASCN is $4$ close to $S_z$ and $-4$ close to $S_{-z}$, while it is $0$ close to $S_x$.
 The material  $\alpha$-BiBr is therefore a SCI for the spin $S_z$ but not for the spin $S_x$. It will have SHC for the $S_z$ spin direction, while it for the $S_x$ direction the SHC will be negligible. In this case the ASCN is a topological invariant of the material. This follows
 from the fact that inversion symmetry is present, it commutes with the spin operators, and 
 it protects the Chern numbers on both positive and negative bands.

   It is important to highlight that the material $\alpha$-BiBr possesses an even more interesting topological structure. It has been found  that 
 both the positive and the negative bands possess a non-trivial Chern-Simons $\theta$-term. \cite{spin-resolved-3d}
 This implies that, independently of the direction of the spin, the surface SHC is non-trivial on this material. This material $\alpha$-BiBr was firstly recognized as a strong topological crystalline insulator \cite{BiBr_Hsu_2019, BiBr_Tang_2019}, and recently has been further
 characterized as T-double axion insulator\cite{spin-resolved-3d}.

\begin{itemize}
\item[5.] {\it If the Chern-Simon axion coupling term is non-zero, and the positive spin projection eigenvectors are mapped to the negative ones, then the ASCN is generically non-zero in all spin directions. In particular, the SHC is non-trivial.}
\end{itemize}
Whenever the Chern-Simons coupling term is non-zero (the Kane-Mele $\mathbb{Z}_2=1$ invariant for time-reversal invariant systems), and the positive spin projection eigenvectors are mapped to the negative ones, then there cannot be a spin gap in any direction of the spin. If there is a spin gap, then the Bloch bundle is a sum of the positive and the negative eigenvectors, and since one is mapped to the other, the Chern-Simons coupling term of the total Bloch bundle is twice the one of the positive eigenvectors. Hence the Chern-Simons coupling term is trivial. This implies that the non-triviality of the Chern-Simons coupling term, together with a symmetry that maps positive to negative eigenvectors (such as time reversal, or a composition of time reversal with a translation), implies the non-triviality of the ASCN.
The non-triviality of the ASCN implies the non-triviality of the SHC.
This can be taken as one explanation of the the non-triviality of the SHC on TIs \cite{Quantized-QSHE,Fukui07}.
 
\begin{itemize}
\item[6.] {\it In strong topological insulators $(\mathbb{Z}_2=1)$ the value of the ASCN $\overline{c}_{1,ab}^{S_n}$
determines the spin Chern numbers on the planes $k_c=0$ and $k_c=\pi$ with $abc$ a base. If $2s < \overline{c}_{1,ab}^{S_c} < 2s+2$ for $s \in \mathbb{Z}$ then the spin Chern numbers $c_{1,ab}^{S_n}(0)$ and $c_{1,ab}^{S_n}(\pi)$  on the planes $k_c=0$ and $k_c=\pi$
are $2s$ and $2s+2$ (not necessarily in this order).}
\end{itemize}
 In strong topological insulators there is only one pair of SWPs with opposite chirality and position.
If the ASCN $c_{1,ab}^{S_n}$ lies between $2s$ and $2s+2$ for $s \in \mathbb{Z}$,  then the spin Chern numbers $c_{1,ab}^{S_n}(l)$ can only take values of $2s$ and $2s+2$. Hence the spin Chern numbers 
$c_{1,ab}^{S_n}(0)$ and $c_{1,ab}^{S_n}(\pi)$ can only be $2s$ and $2s+2$. In the STI Bi$_2$Te$_3$ the ASCN $c_{1,xy}^{S_z}=-3.16$, hence the spin Chern numbers on the planes $k_z=0$ and $k_z=\pi$
can only be $-2$ and $-4$. The calculation shows indeed that $c_{1,xy}^{S_z}(0)=-2$ and $c_{1,xy}^{S_z}(\pi)=-4$.
 
 \vspace{0.2cm}

The ASCN is therefore a good indicator for the existence of topological properties of the spin spectrum. 
It is not in general a topological invariant, in the sense that adiabatic deformations of the
Hamiltonian will not leave the ASCN fixed unless the ASCN is an even integer and there
is a symmetry commuting with the spin operator that protects Chern numbers.

When the ASCN is non-zero, it implies a corresponding non-zero value for the SHC, and the intensity of the SHC is linked to the magnitude of the ASCN. In Table \ref{tablematerials} we have collected the values
of the ASCN and the SHC for different materials with topological properties. It is important to note that
their values are related, and moreover, that we can deduce from the value of the ASCN the fact that
$\alpha$-BiBr has a projected spin gap and therefore it is a spin Chern insulator. 
Additionally, we deduce that the spin Chern numbers for MnBi$_6$Te$_{10}$ are $-6$ and $-4$ on the $k_z=0$ and $k_z=\pi$ planes, respectively.

The ASCN can also be calculated in Weyl semimetals. The occupied states restricted to a plane $k_c=l$ are gapped
except where the energy meets the Fermi energy, which happens only at a finite number of $k$-points.
Moreover, the projected spin operator is always gapped for the planes $k_c=l$ except for a finite number of planes.
Hence the SCN is well defined for all planes $k_c=l$, except for a finite number of planes where Weyl points are located.
In the case of Weyl semi-metals, the value of the ASCN will be very close to the one of the SHC since the eigenvalues of the projected spin operator will be close to $1$ and $-1$\cite{Felser-wp}. 

Note that the {\it average Chern number} (ACN) can also be define on the occupied states:
\begin{equation}
\overline{c}_{1,ab} = \tfrac{1}{2\pi} \int_{0}^{2 \pi} c_{1,ab}(l)  dl.
\end{equation}
Here $c_{1,ab}(l)$ is the first Chern number of the occupied states restricted to the plane $k_c=l$.
In this case it follows that the AHC is a multiple to the ACN, no matter what the Fermi energy level is. In formulas we have
\begin{align}
	\sigma_{ab} = \tfrac{e^2}{h} 2\pi \cdot \overline{c}_{1,ab}.
\end{align}

So we can interpret the AHC as the average contribution of the Chern numbers across parallel planes in reciprocal space. 
This number is therefore directly linked with the distance in momentum space of the Weyl points in Weyl semimetals.  \cite{ahe-in-weyls,AHE-in-WS}.

\section{3D BHZ model} \label{section 3D BHZ model}

We have carried extensive calculations for the ASCN and its relation to the SHC in the 3D BHZ model of four bands \cite{BHZ}. The BHZ Hamiltonian can be written as:
\begin{align}
H({\bf k}) = M({\bf k})\tau_3\sigma_0 + A({\bf k})\tau_1\sigma_3 + C({\bf k})\tau_2\sigma_0 + D({\bf k})\tau_1\sigma_1,
\end{align}
with 
\begin{align}
M({\bf k})=& M_0 -B_0\left(\cos(k_x) + \cos(k_y)+\cos(k_z)\right),\\
A({\bf k}) = & A_0 \sin(k_x), \\
C({\bf k}) = & C_0 \sin (k_y), \\
D({\bf k}) = & D_0 \sin(k_z),
\end{align}
 spin matrices $S_x=\tau_0\sigma_1$, $S_y=\tau_3\sigma_2$ and $S_z=\tau_0\sigma_3$,
time reversal $\mathbb{T}= i \tau_0 \sigma_2 \mathbb{K}$ where $\mathbb{K}$ is complex conjugation, and inversion $I=\tau_0\sigma_3$ \cite{Gonzalez-Uribe}.

Whenever $D_0=0$ we have the original 2D BHZ Hamiltonian consisting of two disconnected blocks related
by time reversal symmetry. In this context, the non-trivial off-diagonal term  $D$ could be interpreted as the
spin-orbit coupling term in the 3D layered system. This new term also breaks the commutativity with the spin operator.

 In matrix form the Hamiltonian is
\begin{equation} 
\label{Hamiltonian}
H({\bf k}) =
\left(\begin{matrix}
M & A+iC & 0 & D \\
(A+iC)^* & -M & D & 0 \\
0 & D &  M  & - (A+iC)^*\\
D & 0 & -(A+iC) &-M 
\end{matrix}\right),
\end{equation}
it preserves time reversal symmetry and inversion since the function $ M({\bf k})$ is even and the functions $A({\bf k}),C({\bf k}),D({\bf k})$ are odd with respect
to inverting ${\bf k}$.

The energy of the Hamiltonian is degenerate with $E= \pm \lambda$, and 
\begin{align}
\lambda = \sqrt{(M^2 + A^2+C^2 + D^2)}.
\end{align}
One choice of eigenvectors of the Hamiltonian is:
\begin{align}
\nu_1 : = & (M-\lambda,A-iC,0,D)^T   \label{nu1}\\
\nu_2 : = & (A+iC, -M-\lambda,D,0)^T   \label{nu2}\\
\nu_3 : = & (M + \lambda, A-iC, 0,  D)^T    \label{nu3}\\
\nu_4 : = & (-(A+iC),M-\lambda, -D, 0)^T \label{nu4}
\end{align}
with $\nu_1$ and $\nu_2$ with energy $-\lambda$ and $\nu_3$ and $\nu_4$ with energy
$\lambda$.

The energy of the spectrum is gapless whenever $M=A=C=D=0$ and this happens
only whenever $\tfrac{M_0}{B_0}=-3,-1,1,3$. Otherwise the energy has a gap.

Furthermore note that $\mathbb{T}I$ preserves the energies and the ${\bf k}$ points.
Therefore $\mathbb{T}I \nu_1$ is also an eigenvector of the Hamiltonian and it is spanned
by $\nu_1$ and $\nu_2$. In the base given above one can calculate explicitly the coefficients of both $\mathbb{T}I \nu_1$ and $\mathbb{T}I \nu_2$ with respect to $\nu_1$ and $\nu_2$. 

More important is the spin operator and its projected version. Consider the projected version $P S_n P$ of the spin operator $S_n$  where $P$ is the projection into the occupied states and $n$ is a generic direction. In the BHZ model we have
\begin{align}
P = | \psi_1 \rangle \langle \psi_1 |+| \psi_2 \rangle \langle  \psi_2|
\end{align}
where the normalized base is
\begin{align}
\psi_i = \tfrac{1}{|\nu_i|}\nu_i.
\end{align}

The projected spin operator acts on the vector spaces of occupied states and returns 
a linear combination of occupied states. The spectrum of the projected spin operator in the reciprocal space will be denoted spin spectrum, and it splits into two main cases. Either the spin spectrum is gapped or not.

In the BHZ Hamiltonian the spin spectrum is not gapped whenever $-3 < \tfrac{M_0}{B_0} <3$, and since the energies are always degenerate, the spin spectrum crosses the 0 line in points. These points are called spin Weyl points (SWP) \cite{Gonzalez-Uribe}.

\subsection{ASCN for $S_z$}

Focusing  at the projected $S_z$ spectrum, we know that the SWPs are located at the points \cite{Gonzalez-Uribe}:
\begin{align}
(0,0, \arccos(\tfrac{M_0}{B_0} + 2 ) )  &\ \ \ \mathrm{for} \ \ -3<\tfrac{M_0}{B_0} <-1 \\
\begin{matrix}
(0,\pi, \arccos(\tfrac{M_0}{B_0}))\\
(\pi,0, \arccos(\tfrac{M_0}{B_0}))
\end{matrix} & \ \ \ \mathrm{for} \  \ \  -1<\tfrac{M_0}{B_0}<1\\
(\pi,\pi,\arccos(\tfrac{M_0}{B_0} -2 ) ) & \ \ \ \mathrm{for} \ \ \ 1<\tfrac{M_0}{B_0}<3 
\end{align}

Denote by ${\bf k}_0$ any of these points. 
Note that on these points, the projection on the valence bands of the $S_z$ operator applied of the valence eigenstates is precisely zero.
What this means theoretically is that on ${\bf k}_0$
\begin{align}
{S}_z& \psi_1({\bf k}_0) \rangle =
 \sum_{j=3,4} |\psi_j({\bf k}_0) \rangle \langle \psi_j({\bf k}_0) | \widehat{S}_z \psi_1({\bf k}_0) \rangle.
\end{align}
and therefore the $S_z$ of the valence bands on the point ${\bf k}_0$ is all conduction (see Fig. \ref{figure_spin_projection}). This means that the 3D BHZ models an insulator for the Hamiltonian but not for the projected spin.

Around the SWP we could put a small 2D sphere surrounding it. On this sphere
the $S_z$ operator splits into valence and conduction part. The way the negative projected spin spectrum eigenvectors twirl around the spin Weyl point is measured by the chirality of the point (Chern number). The SWPs are topologically protected whenever their chiralities are different from zero. Their positions may vary through adiabatic perturbations of the Hamiltonian, but their chiralities will remain fixed.

The values of the SCN for the projected spin $S_z$ operator on the planes $k_j =0, \pi$ for $j=x,y,z$ permits us to determine the ASCN  on the three cases. That is:
\begin{align}
\overline{c}_{1,xy}^{S_z} = \left\{
\begin{matrix}
-2+\tfrac{2}{{\pi}} \arccos(\tfrac{M_0}{B_0} + 2) &  \mathrm{for}  & -3<\tfrac{M_0}{B_0}<-1  \\
 2 - \tfrac{4}{ \pi} \arccos(\frac{M_0}{B_0})& \mathrm{for} & -1<\tfrac{M_0}{B_0}<1   \\
 \tfrac{2}{{\pi}}  \arccos\ ( \tfrac{M_0}{B_0} - 2) &  \mathrm{for}  & 1<\tfrac{M_0}{B_0}<3
\end{matrix}
\right.
\label{shcequa}
\end{align}

These results are highlighted by the red line depicted in Fig. \ref{shcvsmobo}.  Here is important to notice that non-integer values of the ASCN implies the existence of SWPs.

The SOC could be interpreted as the matrix $D\tau_1\sigma_1$ and its intensity as the value of $D_0$. 
In Fig. \ref{shcvsmobo} a) the ASCN $\overline{c}_{1,xy}^{S_z}$ has been plotted with respect to the value of $\tfrac{M_0}{B_0}$, together
with the SHC $\sigma_{xy}^z$ for different values of $D_0$. 

It is important to note that the ASCN remains unaffected by the value of $D_0$, serving as a direct measure of the SWP's distance.
Notably, in the case of  $D_0=0$, the ASCN aligns with the SHC and the intensity of the SHC decreases for bigger $D_0$;
however, both the ASCN and the SHC retain non-zero values simultaneously.

\subsection{SWP for generic spin} \label{subsection SWP for generic spin}

Consider the unitary vector $n=(\alpha,\beta,\gamma)$ and the generic spin $S_n= n \cdot (S_x,S_y,S_z)$
 
If $\psi_j$ are the eigenvectors of the Hamiltonian forming a unitary base (norm one and
perpendicular to one another), then the projected spin matrix is defined as follows:
\begin{align}
(M_{{S}_n})_{ij}= \langle \psi_i|S_n|\psi_j \rangle\ \ i,j \in \{1,2\}.
\end{align}

The projected spin eigenvalues vanish whenever the whole projected spin matrix
matrix vanishes. Note that in this case we could use the degenerate basis $\{\nu_1,\nu_2\}$ 
of Eqns. \eqref{nu1} and \eqref{nu2} in order to solve the equations
\begin{align}
 \langle \nu_i|S_n|\nu_j \rangle\ = 0 \ \  \ i,j \in \{1,2\}.
\end{align}
Let us calculate first the matrix elements $\langle \nu_i |S_{x_k}|\nu_j \rangle$
for $k=1,2,3$:
\begin{align}
 \langle \nu_1 |S_x|\nu_1 \rangle\   &=  2DA \\
 \langle \nu_1 |S_x|\nu_2 \rangle\   &= -2\lambda D \\
\langle \nu_2 |S_x|\nu_1 \rangle\   &= -2\lambda D\\
\langle \nu_2 |S_x|\nu_2 \rangle\   &= 2DA
\end{align}
\begin{align}
 \langle \nu_1 |S_y|\nu_1 \rangle\   &=  -2DC \\
 \langle \nu_1 |S_y|\nu_2 \rangle\   &= 2i\lambda D \\
\langle \nu_2 |S_y|\nu_1 \rangle\   &= -2i\lambda D\\
\langle \nu_2 |S_y|\nu_2 \rangle\   &= -2DC
\end{align}
\begin{align}
 \langle \nu_1 |S_z|\nu_1 \rangle\   &=  (M - \lambda)^2 + A^2+C^2 - D^2 \\
 \langle \nu_1 |S_z|\nu_2 \rangle\   &= 2 (A+iC) \lambda \\
\langle \nu_2 |S_z|\nu_1 \rangle\   &= 2 (A-iC) \lambda\\
\langle \nu_2 |S_z|\nu_2 \rangle\   &= (M + \lambda)^2 + A^2+C^2 - D^2
\end{align}

The projected spin matrix has for entries:
\begin{align}
(M_{{S}_n})_{11}= &  2\alpha DA -2\beta DC  \\
 &+ \gamma( (M - \lambda)^2 + A^2+C^2 - D^2) \nonumber\\
(M_{{S}_n})_{21}= &  -2 \alpha \lambda D + 2 i \beta \lambda D +2\gamma \lambda (A+iC)\\
(M_{{S}_n})_{12}= &  -2 \alpha \lambda D - 2 i \beta \lambda D +2\gamma \lambda (A-iC)\\
(M_{{S}_n})_{22}= &  2\alpha DA -2\beta DC\\
 &+ \gamma( (M + \lambda)^2 + A^2+C^2 - D^2) 
\nonumber
\end{align}

The projected spin matrix has zero eigenvalues whenever all entries are zero. Subtracting
$(M_{{S}_n})_{11}=0$ from $(M_{{S}_n})_{22}=0$ we see that $M=0$, and therefore
\begin{align} \label{eqn M11}
 \alpha  DA - \beta DC + 2 \gamma (A^2+C^2) =0.
\end{align}
From
$(M_{{S}_n})_{21}=0$ we see that 
\begin{align}
 \alpha D  -i \beta D + \gamma (A+iC) =0.
\end{align}

Replacing $A+iC= \frac{-\alpha + i \beta}{\gamma }D$ in Eqn. \eqref{eqn M11} we see that the 
equation is satisfied. Then we only need to consider the equations $M=0$ and 
$\alpha D  -i \beta D + \gamma (A+iC) =0$. These equations can be written as:
\begin{align}
\cos(k_x)+ \cos(k_y) + \cos(k_z) &= \tfrac{M_0}{B_0} \\
\gamma A_0 \sin(k_x) + \alpha D_0 \sin(k_z) &=0\\
\gamma C_0 \sin(k_y) - \beta D_0 \sin(k_z) &=0.
\end{align}

Let us see some explicit choices of $n$:
\begin{itemize}
\item $n=(0,0,1)$, in this case we have that $k_x,k_y=0,\pi$ and $k_z = \arccos(\tfrac{M_0}{B_0} + \{2,0,-2\})$ depending on the location of $\tfrac{M_0}{B_0}$.
\item $n=(1,0,0)$, $n=(0,1,0)$ or any combination of the two. In this case the system is degenerate and the matrix is zero on the whole $k_z=0$ plane.
\item $n=(\tfrac{1}{\sqrt{2}},0, \tfrac{1}{\sqrt{2}})$, and moreover $A_0=C_0=D_0$, then we have that
$k_y=0,\pi$, $k_x=-k_z$ and $2\cos(k_x)= \tfrac{M_0}{B_0} \pm 1$.
\end{itemize}

In general, for any direction $n$ of the spin, there is a pair of opposite chiralities SWPs in the Bulk
whenever $1<|\tfrac{M_0}{B_0}|<3$. For $|\tfrac{M_0}{B_0}|<1$ there are two pairs of opposite chiralities SWPs. Their location is defined by the solution of the equations defined above.

In Fig. \ref{shcvsmobo} b) we have plotted the ASCN $\overline{c}_{1,xy}^{S_n}$ and the SHC for values of $n=(\sin(\theta), 0, \cos(\theta))$ where $0 \leq \theta \leq \pi$. The ASCN only depends on the location and chiralities of the SWPs
for each choice of $n$, while the SHC can be calculated by the expression
\begin{align}
\sigma_{xy}^n = \sin(\theta) \sigma_{xy}^x +\cos(\theta) \sigma_{xy}^z.
 \end{align}
 
In Figure \ref{shcvsmobo} b) it is observed that the SHC exhibits alignment with the ASCN when the spin direction rotates along the $k_y$ axis. 
In this graph, it is noted that both ASCN and SHC undergo a change in sign from $S_z$ (angle=0) to $S_{-z}$ (angle=$\pi$).

\begin{figure}
	\includegraphics[width=8.9cm]{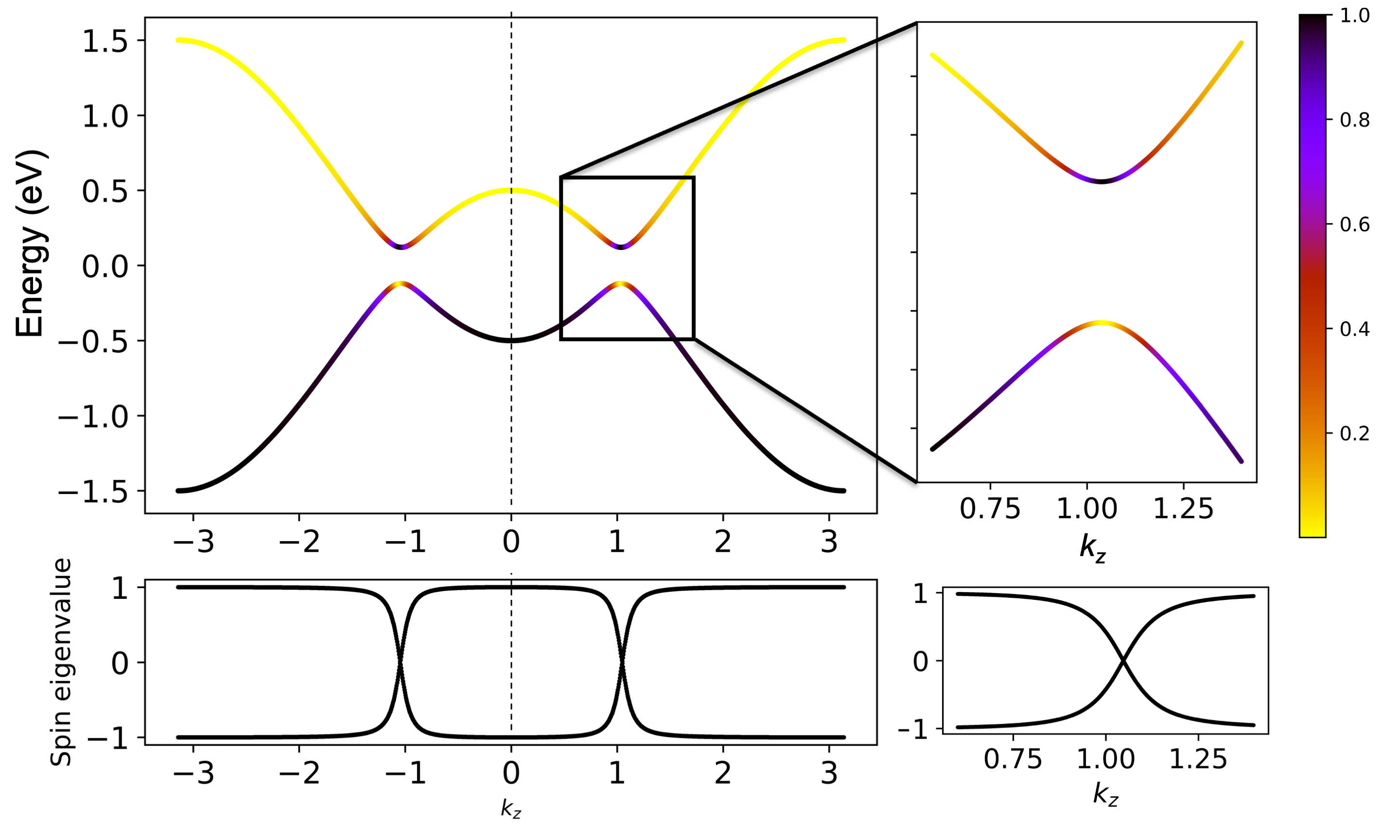}
	\caption{Graph of the projection of the positive projected spin $S_z$ eigenvector into the valence and conduction bands
on the 3D BHZ model at the $k_z$ axis on the point $(k_x,k_y)=(0,\pi)$ for $\tfrac{M_0}{B_0}= 0.5 $, $A_0=C_0=1$, $D_0=0.11$. The SWPs are located on this axis and the graph of the projected spin eigenvalues lies at the bottom.
The top graphs are the degenerate energies of the four bands, and the color represents the square of the norm
of the projection of the positive spin eigenvector of $PS_zP$  into the valence and the conduction states respectively. If $\phi$ is the positive eigenvector of $PS_zP$, then the color on the upper bands represent the value
$|\langle \psi_3 |\phi \rangle |^2 + |\langle \psi_4 |\phi \rangle |^2$ and on the bottom bands the value
$|\langle \psi_1 |\phi \rangle |^2 + |\langle \psi_2 |\phi \rangle |^2$. In the location of the SWPs, the 
spin $S_z$ of both the valence states $\psi_1$ and $\psi_2$, is all conduction. 
Even though the 3D BHZ model has an energy gap, at some points in momentum space
the spin of the valence states lies on the conduction band. The BHZ models an energy insulator which
 does not insulate the spin.
 }
	\label{figure_spin_projection}
\end{figure}  

In Fig. \ref{figure_spin_projection} we have plotted the energy bands on the $k_z$-axis for $(k_x,k_y)=(0,\pi)$ 
 for specific choice of constants. The SWPs are located on this axis. We have colored the energy bands
with the square of the norm of the projection of $S_z |\phi \rangle$ on the valence and the conduction bands
where $\phi$ is the eigenvector of the projected spin operator $PS_zP$ with smallest positive eigenvalue.
Note that at the SWPs, the spin of the valence energy states is all conduction. 
Namely, even though the BHZ models an insulator in terms of its energy spectrum, its is not an insulator on the spin spectrum.

\section{Materials}

\begin{figure*}
	\includegraphics[width=18cm]{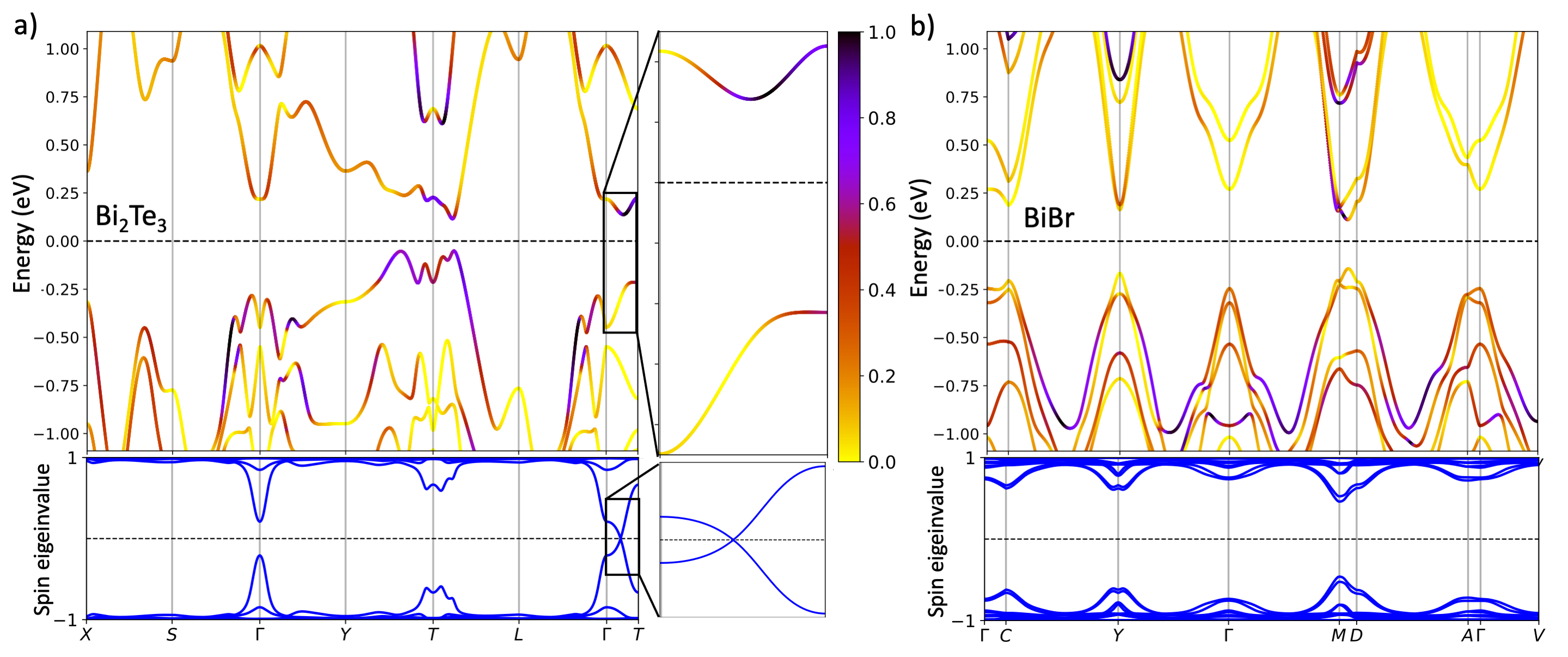}
	\caption{Energy bands and projected spin $S_z$ eigenvalues for the strong topological insulator Bi$_2$Te$_3$ and the weak topological insulator $\alpha$-BiBr.
		The color on the energy bands represent the square of the norm of the projection of the smallest positive projected spin eigenvector into each pair of degenerate bands, both valence and conduction. If $\phi$ is the eigenvector of the smallest positive eigenvalue of $PS_zP$, the color on each pair of consecutive degenerate bands $\psi_{2n-1}, \psi_{2n}$ is given by the value
		$|\langle \psi_{2n-1} | S_z \phi |\rangle|^2 +|\langle \psi_{2n} | S_z | \phi \rangle|^2$.  a) Bi$_2$Te$_3$ is a strong topological insulator, and as such, it has SWPs on the T-$\Gamma$-T path.  The chiralities of these SWPs is $\pm 1$ and the ASCN is $c_{1,xy}^{S_z}=3.16$, which means that the SWPs are located almost in the middle of the T-$\Gamma$ path. Note that the value of the intrinsic SHC is $\sigma_{yx}^z=3.27$ which
		is very close to the ASCN. The eigenvector of the projected spin operator
		with smallest positive eigenvalue is entirely conduction on the SWPs. This can
		be seen in the $\Gamma$-T path.  b) $\alpha$-BiBr is a spin Chern insulator since the value of the ASCN  $c_{1,xy}^{S_z}=-4$ is a non-zero even integer. The spin insulation can be read from the fact that the ASCN is an even integer, and the topological feature is given by the constant value of $-4$ for the SCN on the $k_xk_y$ planes. Since this SCN is a multiple of $4$, the Chern number of the positive and negative projected spin eigenvectors is equal to $2$ and $-2$ respectively. Hence the number of pairs of negative eigenvalues of the inversion operator is even and therefore the strong topological insulator marker $\mathbb{Z}_2$ of Fu-Kane-Mele is equal to zero. Nevertheless, it has been
shown \cite{spin-resolved-3d} that the Chern-Simons axion $\theta$-term of both the positive and the negative bands is non-trivial. Thus, independent of the spin direction,
the surface SHC contribution is non-trivial. This material $\alpha$-BiBr was firstly recognized as a strong topological crystalline insulator \cite{BiBr_Hsu_2019, BiBr_Tang_2019}, and
recently has been characterized also as a $\mathbb{T}$-double axion insulator\cite{spin-resolved-3d}.
		}
	\label{materials}
\end{figure*}

In order to demonstrate the applicability of the current methodology to real-world materials,
we have calculated the ASCN and SHC for different insulator materials.
Here we contrast the results for the case of Bi$_2$Te$_3$, MnBi$_2$Te$_4$,  
MnBi$_6$Te$_{10}$ and $\alpha$-BiBr, which were chosen due to their predicted topological properties \cite{Topological-quantum-chemistry,Magnetictopologicalquantumchemistry,MnBi6Te10,BiBr_Hsu_2019,BiBr_Tang_2019,spin-resolved-3d}.
Previous works classified them as either strong topological insulators, axion insulators, or quantum Hall spin insulators. 
The structural parameters for these materials were obtained from the Materials Project\cite{materialsproject} and previous studies\cite{MnBi6Te10}.

The calculations were performed using a combination of first-principles methods based on the VASP \cite{vasp}, the Wannier90 \cite{wannier90}, and the Pythtb code \cite{pythtb}, with detailed computational setups outlined in our previous work\cite{Gonzalez-Uribe}. However, computational tools have recently been created to study the spin topology of materials \cite{spin-resolved-3d,BerryEasy_2024} 

We evaluated the electronic and spin spectra of these materials, including the projection of spin projected eigenvalues ($PS_zP$) onto the electronic bands, as depicted in Fig. \ref{materials}.
For the case of Bi$_2$Te$_3$, we observed a pronounced concentration of valence spin information around a singular point in the conduction band along the $\Gamma$T $k$-path. 
This localization of spin information suggests the existence of a spin Weyl point along the $\Gamma$T path, as depicted in Fig. \ref{materials} a).
Remarkably, the projected spin spectrum calculation unveils the presence of spin Weyl points both in the T$\Gamma$ and -T$\Gamma$ paths, which represent the diagonal direction in the rhombohedral phase or $z$ axis in cartesian coordinates.
Consequently, the ASCN is expected to be proportional to the distance between these SWPs, as in the 3D BHZ model. 
Indeed, we have found that the both ASCN and SHC exhibit a proportional relationship with the SWPs distance. 
Table \ref{tablematerials} presents the value of ASCN and SHC for Bi$_2$Te$_3$, demonstrating a topological response across all $k_z$-planes in the hexagonal BZ. 
Notably, the SHC signal is particularly pronounced along the $z$-axis, coinciding with the location of SWP in reciprocal space.
This result aligns with the layered structure of Bi$_2$Te$_3$, where the $z$ planes correspond to the layers in real space \cite{Zhang-bi2se3}.

 \begin{table}[htbp]
	\centering
	\caption{Average Spin Chern Number ($\overline{c}^{z}_{1,ij}$) and Spin Hall Conductance ($\sigma^{z}_{ij}$ ) values for topological materials Bi$_2$Te$_3$, MnBi$_2$Te$_4$ and MnBi$_6$Te$_{10}$  and $\alpha$-BiBr, where MnBi$_2$Te$_4$ and MnBi$_6$Te$_{10}$  are taken in their antiferromagnetic phase. The table includes ASCN and SHC values are given in units of $-e\pi$. The last row classifies the materials as strong topological insulator (STI), axion insulator (AI) or topological crystalline insulator \cite{BiBr_Hsu_2019, BiBr_Tang_2019} (TCI).}
	\label{tablematerials}
	\begin{tabular}{ccccc}
		\hline
		\hline
		&Bi$_2$Te$_3$ & MnBi$_2$Te$_4$ & MnBi$_6$Te$_{10}$ & $\alpha$-BiBr \\
		\hline
		$\sigma^z_{yx}$                    &   $3.27$       & $-2.70$ & $-4.48$ &  $-3.96$ \\
		$\overline{c}^{S_z}_{1,xy}$ &  $3.16 $     & $-2.60 $&$ -5.04$ & $-4$ \\
		$\sigma^z_{xy}$                     & $-3.25$    &  $2.70$  & $4.48$ & $3.66$  \\
		\hline
		Type & STI       &  AI &  AI	& TCI \\
		\hline
		\hline
	\end{tabular}
\end{table}

We have performed an analysis of the electronic and projected spin spectrum of $\alpha$-BiBr, uncovering a nonzero energy and projected spin gap throughout the entire Brillouin zone. 
This characteristic serves to classify the $\alpha$-BiBr as both as a material with both energy gap and spin gap, as it shown in Fig. \ref {materials} b), in agreement with Lin \textit{et al} \cite{spin-resolved-3d}. 
Furthermore, no significant exchange of spin information between the conduction and valence bands is evident.
However, we have observed a constant $\overline{c}^{z}_{1,xy}$ value of $-4$ along all the $k_z$ planes and zero of the $k_x$ and $k_y$ planes in the Brillouin zone for $\alpha$-BiBr, 
which is consistent with the $\sigma^z_{yx}\sim -4$ as presented in Table \ref{tablematerials}. 
The alignment between SHC and ASCN reflects the full topological response of $\alpha$-BiBr perpendicular to the $z$-axis. 
This result is hidden for the $\mathbb{Z}_2$=0 index for this material, highlighting the efficacy of ASCN as a tool for extracting valuable insights into the spin response of topological insulators. Here it is worth mentioning that each group of bands, the positive and the negative, possess non-trivial Chern-Simons $\theta$-term\cite{spin-resolved-3d}. This implies that independent of the spin direction, it will give rise to a surface SHC contribution in a finite sample.

It is found that $\alpha$-BiBr displays a vanishing ASCN for the $S_{x}$ and $S_{y}$ spin components, suggesting a lack of topological response in SHC for these spin directions. 
Our calculations confirm $\alpha$-BiBr as a 3D quantum spin Hall insulator ($S_{z}$). 
The observed anisotropy between $\sigma^z_{xy}$ and $\sigma^z_{yx}$ responses is further supported by symmetry analysis of the SHC tensor, indicating distinct components within this space group (\#12).
This convergence of ASCN and SHC values highlights the unique topological nature of $\alpha$-BiBr.

Regarding the axion insulators (AI) MnBi$_2$Te$_4$ and MnBi$_6$Te$_{10}$, we also see that
the values of the ASCN and SHC are very similar. In these two materials time reversal coupled with a translation is a preserved symmetry, and therefore the Chern numbers of the positive and negative 
projected spin eigenvectors are inverse to one another. Therefore we can conclude that
the spin Chern number on all $k_z$ planes for MnBi$_2$Te$_4$ are less than $-2$ and for
MnBi$_4$Te$_{10}$  are less than $-4$. 
These findings align with the non-trivial results of the pseudo SCN for the MnBi$_2$Te$_4$ family, as obtained by Wang \textit{et al} \cite{pseudoSCN}.

With the SHC alone we may not distinguish the materials in Table \ref{tablematerials}, but with the
incorporation of the ASCN we differentiate $\alpha$-BiBr from the others.
The incorporation of the ASCN into the set of material classifiers will help distinguish QSHIs
from strong topological and axion insulators, and moreover, it will permit to distinguish the topological 
features underlying the projected spin operator.

It is also important to remark that the ASCN calculations are significantly less intensive compared to SHC calculations. The former needs 3D dense $k$-grids while the latter only needs a dense $k$-grid in one dimension.
The ASCN can also be calculated in collinear ferromagnetic or antiferromagnetic when the spin $z$ component can be consider a good quantum number in the weak SOC limit.
These results indicate the practical advantage of employing ASCN as a computational tool and topological indicator, in scenarios where the efficient calculation of SHC is limited.

\section{Conclusion}

  The ASCN consists of a family of numbers associated with a choice of spin direction
and a choice of plane of integration. These numbers provide hindsight of the
spin properties of an insulator material. If the ASCN is non-zero for a specific choice of spin direction, then a signal in the SHC should be expected. 
Moreover, if the material is an STI, then
the ASCN is different from zero. Therefore, materials with non-zero values for the ASCN 
include all STIs. 
When the ASCN is an even integer, then there is a gap in the projected spin operator, and therefore the Kane-Mele invariant $\mathbb{Z}_2$ is trivial. Moreover, if there
is a symmetry commuting with the spin which protects Chern numbers, the value of the ASCN is a topological invariant. \\
From a computational perspective, calculating the ASCN is less computationally intensive compared to the SHC.
For the latter, a 3D dense grid is mandatory, while for the former just a dense grid in the direction of integration is necessary. The calculation of the Chern numbers does not require dense grids.
Understanding the ASCN for various spin directions and planes contributes to the comprehension and categorization of insulating materials, enhancing our knowledge of their topological properties and aiding in their classification.

 \vspace{0.4cm}

\section*{Acknowledgments}
The first author gratefully acknowledges the computational time granted on the supercomputer MOGON 2 at Johannes Gutenberg University Mainz (hpc.uni-mainz.de). Additionally, the support from the Universidad Nacional de Colombia (QUIPU code 202010042199) and from MinCiencias through Convocatoria 937 for Fundamental Research is also deeply appreciated.
The second author acknowledges the support of CONACYT through project CB-2017-2018-A1-S-30345-F-3125 and of the Max Planck Institute for Mathematics in Bonn, Germany. 
Both authors thank the continuous support of the Alexander Von Humboldt Foundation, Germany.

\bibliographystyle{naturemag}
\bibliography{Topological}

\end{document}